\documentclass[runningheads]{llncs}
\usepackage{graphicx}
\begin{document}
\title{Leveraging a Federation of Knowledge Graphs to Improve Faceted Search in Digital Libraries}
\titlerunning{Leveraging Knowledge Graphs in Faceted Search}

\author{Golsa Heidari\inst{1}\orcidID{0000-0002-5398-7086} \and
Ahmad Ramadan\inst{1}\orcidID{0000-0002-3238-4315} \and
Markus Stocker\inst{1, 2}\orcidID{0000-0001-5492-3212} \and
S\"oren Auer\inst{1, 2}\orcidID{0000-0002-0698-2864}}
\authorrunning{G. Heidari et al.}
\institute{L3S Research Center \& Leibniz University of Hannover, Germany\\
\email{\{golsa.heidari, ramadan\}@stud.uni-hannover.de}\\
\url{https://www.uni-hannover.de/} \and
{TIB Leibniz Information Centre for Science and Technology, Germany}\\
\email{\{markus.stocker,auer\}@tib.eu} \\
\url{https://www.tib.eu/}}

\maketitle              
%
%

\begin{abstract}
Scientists always look for the most accurate and relevant answers to their queries in the literature. Traditional scholarly digital libraries list documents in search results, and therefore are unable to provide precise answers to search queries. In other words, search in digital libraries is metadata search and, if available, full-text search. We present a methodology for improving a faceted search system on \emph{structured content} by leveraging a federation of scholarly knowledge graphs. We implemented the methodology on top of a scholarly knowledge graph. This search system can leverage content from third-party knowledge graphs to improve the exploration of scholarly content. A novelty of our approach is that we use dynamic facets on diverse data types, meaning that facets can change according to the user query. The user can also adjust the granularity of dynamic facets. An additional novelty is that we leverage third-party knowledge graphs to improve exploring scholarly knowledge.
\keywords{Knowledge Graph \and Scholarly Knowledge \and Information Retrieval \and Search System \and Faceted Search \and Digital Libraries.}
\end{abstract}

\section{Introduction}
A knowledge graph (KG) is a knowledge base that uses a graph-structured data model or topology to combine data~\cite{ehrlinger2016towards}. Knowledge graphs are often used to store interlinked information about entities with free-form semantics. In recent years, knowledge graphs have been presented and made publicly available in the scholarly field, in particular bibliographic metadata including information about entities such as publications, authors, and venues~\cite{farberenhancing}. 

Scholarly Knowledge graphs are knowledge bases for representing scholarly knowledge~\cite{Jaradeh2019}. If scholarly knowledge graphs represent the key content published in papers about the addressed research problem, employed materials, methods, and obtained results, then accurate information can be retrieved from such graphs to satisfy user queries and questions.
Due to the rise of knowledge graph usage among scientists, it is predictable that researchers' method of searching and exploring data is moving in that direction over the next few decades~\cite{hoffman2018smart,jaradeh2019open}.

One of the essential applications of scholarly knowledge relies on data retrieval. Various search systems are implemented to help scientists for exploration of accurate data. An example of that is faceted search. Faceted search is a high-efficiency search method with various applications. Faceted search is a method that augments traditional search systems with a faceted exploration system, allowing users to narrow down search results by applying multiple filters based on the classification of the properties~\cite{Feddoul2019}. A faceted classification system lists each knowledge component along various dimensions, called facets, facilitating the classifications to be reached and managed in multiple forms. Faceted search is widely implemented on bibliographic metadata. However, on data, i.e. the actual content of a paper, it simply cannot be implemented because this data is not structured properly. 

Facets are defined in two categories: Static Facets and Dynamic Facets~\cite{mihindukulasooriya2020dynamic}. Facets in which the values for a facet are taken from a list of predefined values are called static facets. Static facets are useful for categories such as \textit{resource type} that have a limited number of possible values~\cite{zheng2013survey}. In contrast, dynamic facets in which the values for each facet category are derived from the values stored in the knowledge graph are flexible~\cite{feddoul2019automatic}. Once the system determines which values to display for each category, it will show the matching items accordingly. This means that facets are not fixed and will be defined while search~\cite{basu2008minimum}.

The rest of the paper is organized as follows: Section 2 describes the background and related work; Section 3 illustrates our methodology, proposed conceptual model and workflow for improving dynamic faceted search to explore data in federated knowledge graphs; Section 4 describes our implementation of the conceptual model in ORKG\footnote{Open Research Knowledge Graph}; In Section 5 we discuss our work and challenges that we faced; In section 6 we propose some directions for future work; Finally, in Section 7 we conclude the work with a glance to the future work.

\section{Related work}
\subsubsection{Search Systems.}Nowadays, many databases contribute scholarly knowledge such as papers. Although faceted search is exceptionally beneficial for knowledge retrieval, search engines have used it almost at the level of metadata for the scholarly literature. In some disciplines, people also described content in articles in a structured manner and they have built search systems, but their work is limited to one research field.

Google Scholar\footnote{\url{https://scholar.google.com/}} is a well-known example that renders a huge number of results fast and most results are not precise to the user information need. Although it has a vast database, static facets are just defined on the publishing date and, thus, limited support for refining queries. Furthermore, it does not search the content of a paper. Solely a full-text search on the abstract part of a paper when the full text is available. 

Publishers such as IEEE\footnote{\url{https://www.ieee.org/}} and Springer\footnote{\url{https://www.springer.com}} show better results via their search system. Their search results are more accurate and using facets they can limit a huge number of unwanted papers to a more relevant set. But there are still limitations to their search system. The most prominent is that their database is limited to their publications. Therefore a large number of results would be missed. Moreover, while they offer faceted search, their facets are static and identical for all queries.

TIB portal\footnote{\url{https://www.tib.eu/de/}} is a meta catalogue, so it provides more relevant answers to the search. Hence, the results would be more accurate. But the problem of the static facets, however, exists there. 

\subsubsection{Research on Knowledge Graphs and Search Systems.}Most of the scientific discoveries depend on searching and re-using the results of former researchers. Although data and metadata of publications always have been available easily, exploring content of a paper remained inaccessible. Scientists tried to explore how developments in web technology might support that method by implementing semantic improvements to journal articles. 

S. Fathalla et al. claim that research contributions must be transparent and comparable. They designated surveys for research ﬁelds in a semantic way and introduced a knowledge graph that defines the specific research problems, approaches, implementations and evaluations in a structured and comparable way. They offered an ontology to capture the content of survey papers~\cite{fathalla2017towards}. D. Poole et al. worked on semantic science. They focused on having machine-accessible scientiﬁc theories that can be used in making data comparable~\cite{poole2006semantic}.

Some researchers extend the current concept of nanopublications — small items of scientiﬁc results in RDF description — to expand their application range. Nanopublications have been introduced to make it more findable~\cite{kuhn2013broadening,mons2009nano}.

Y. Tzitzikas et al. introduced features and standards for surveying the products in the area of browsing and exploring RDF/S data sets. They introduced information requirements and structures. They provided a generalization of the main faceted exploration/browsing approaches using a small model including states and transitions between states~\cite{tzitzikas2017faceted}. 

Some researchers provide theoretical foundations for faceted search in the context of RDF-based knowledge graphs enhanced with OWL ontology~\cite{arenas2016faceted}. Others in addition to faceted search implementation, proposed a ranking system to order facets, and filtered the answer size to avoid numerous answers on statistical properties of their data set, as well~\cite{manioudakis2020faceted}. 

Shotton et al. published downloadable spreadsheets containing data from within tables and figures and enriched them with information from other articles. They published machine-readable RDF\footnote{Resource Description Framework} metadata both about the article and about the references it cites~\cite{shotton2009adventures}.

LINDASearch presents a middle ware structure to produce information about some of the Open Linked Data Projects such as DBpedia, GeoNames, LinkedGeoData, FOAF proﬁles, Global Health Observatory, Linked Movie Database (LinkedMDB) and World Bank Linked Data~\cite{sanchez2020lindasearch}.

The next section briefly describes how implementing a faceted search over scholarly knowledge supports granular refinement of search queries and would leverage federated knowledge graphs.

\section{Methodology}

\begin{figure}
\includegraphics[width=\textwidth]{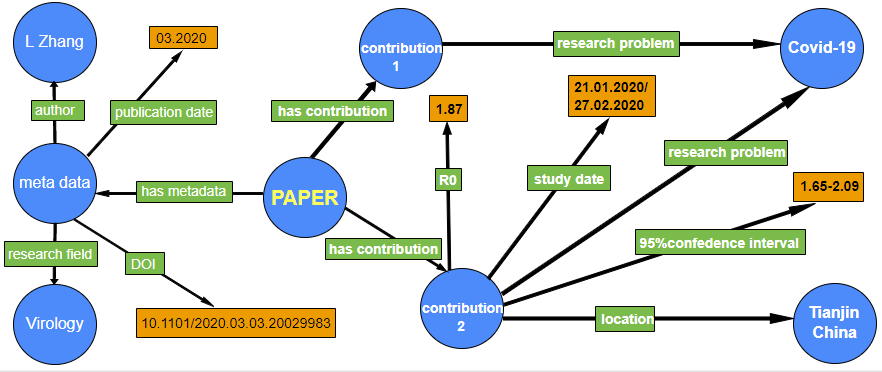}
\caption{An example of semantically representing some of the information contained in a paper in a scholarly knowledge graph.} \label{paper-data.PNG}
\end{figure}

The main idea is to work on different data types to leverage faceted search systems on knowledge graphs. The scholarly knowledge graph which is used for the infrastructure of the faceted search system should not only contain the metadata of the publications, but also semantic, machine-readable descriptions of scholarly knowledge~\cite{oelen2020creating}. Therefore, the knowledge graph would represent some of the content of a publication in a structured manner using inter-linked properties i.e., \textit{study date, study location, method, approaches, research problem}, etc. Figure~\ref{paper-data.PNG} shows how some of the information contained in a scholarly article would be defined in a scholarly knowledge graph.

\subsection{Exploratory Search}

Our search system not only explores the exact data indicated in a paper but also processes some data to narrow down the search results by defining innovative facets. We treat each data type differently. For string data (i.e., properties that have strings for values), a user can select one or more values among all. This is also supported by an auto-complete feature to suggest candidate options. For properties such as \textit{method} and \textit{approach}, all methods used in the papers and all approaches related to them are proposed to the user and can be filtered. For numerical data, users may not only want to filter data by a distinct value but also by a range. Hence, different operators can be selected for the filtering process, specifically greater or smaller than a specific amount. Furthermore, a user can exclude values or even filter data for an interval. Similarly, operators can be applied for values of type date. In addition to including or excluding a date, a duration of a study can be selected as a valid filtering criterion. A date picker is activated on date properties so a user can easily select the date on a calendar. 

In order to have smarter facets to better filter the search results for some data types, we need other knowledge graphs' data. Here is the point that exploration will flow from one knowledge graph to another one. For taxonomic data such as location, we search for the hierarchy in a related knowledge graph. Using API, a third-party knowledge graph can be explored to find the hierarchy of that location. Getting the hierarchy, exploration at various levels of a taxonomy can be done. In other words, different levels of facets will define.

\subsection{Defining Facets}
Facets are defined not only on the metadata of a paper but also on the data, which is essential for each publication. Since facets are defined according to the semantic contribution descriptions for each paper, they are not static and would differ for each query. They are defined dynamically according to the query, and their granularity level can be chosen by the user while querying. For instance, looking for a paper about Covid-19, one would find R0\footnote{The basic reproduction number (R0) is the average number of infections produced by a single infectious person in a population with no immunity.} amounts as a facet. Such facet would not appear when searching mathematics research contributions.
As our focus is on approaching a high-quality search on taxonomic data, these facets are defined in various granularity levels. For instance, \textit{Location} can be explored at the continent level, region level, country level, city level, or even a compound level. 

Our system is supporting such dynamic facets, which are inferred automatically from the respective data types and values. Facets can be different for each query, in contrast to other search systems which use just a predefined set of static facets. 

\section{Implementation}
\begin{figure}
\includegraphics[width=\textwidth]{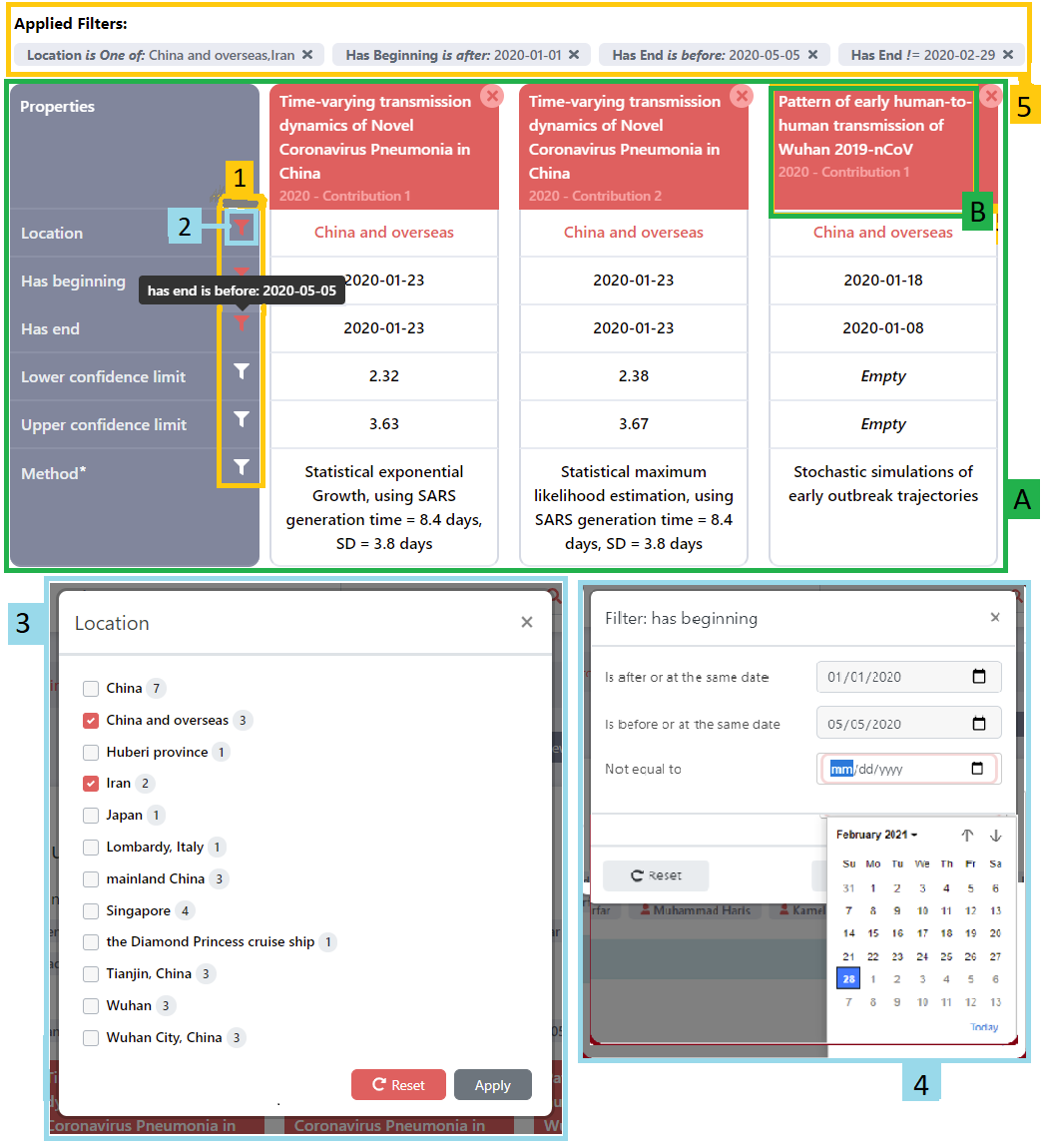}
\caption{Comparison and faceted search UI showing a comparison of studies on the COVID-19 reproductive number estimates and corresponding scholarly knowledge managed by the ORKG. The upper part (A) highlights the tabular comparison of the individual contribution descriptions (B) extracted from scientific papers that employ knowledge graph properties shown on the left-most side of box A. Numbered bounding boxes illustrate the search facilities that are available to users. 1) Filter icons to select the value for the properties. 2) Upon clicking a filter icon certain dialogue boxes like (3) or (4) appear. 3) A selection prompt of location candidate facets. 4) A selection prompt of study date facets. Different facet types call for different selection options. 5) Currently activated filters on the comparison.}
\label{comparison-facets.png}
\end{figure}

The Open Research Knowledge Graph (ORKG)\footnote{\url{https://www.orkg.org/orkg/}} is an online resource that semantically represents research \textit{contributions} (from papers) in the form of an interconnected knowledge graph~\cite{oelen2020creating}. It provides machine-actionable access to scholarly literature that habitually is written in prose~\cite{fathalla2017towards}, and enables the generation of tabular representations of contributions as \textit{comparisons}. Given described papers and their research contributions, it is possible to compare the contributions addressing a specific problem, across the scholarly literature. Figure~\ref{comparison-facets.png} shows a comparison in ORKG. We implemented our faceted search system for ORKG comparisons.

Some research contribution descriptions in the ORKG are specified by predefined templates. These templates support the dynamic and automated construction of facets for ORKG comparisons. Facets are defined on the different properties in a comparison. 

\begin{figure}
\includegraphics[width=\textwidth]{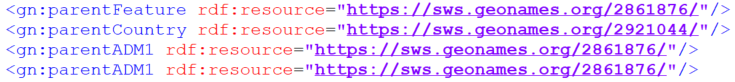}
\caption{A subset of RDF triples from the Geonames representation of the \textit{City of Bonn} entity indicating the \textit{parent feature} as well. } \label{rdf.png}
\end{figure}

\begin{figure}
\includegraphics[width=\textwidth]{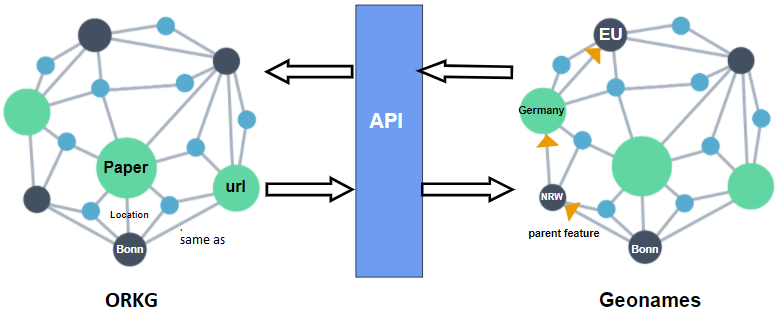}
\caption{The workflow between the ORKG and Geonames knowledge graphs.} \label{geonames.png}
\end{figure}

In order to illustrate how we can leverage other knowledge graphs, we use Geonames\footnote{\url{https://www.geonames.org}} for the \textit{Location} property. Each instance of the \textit{Location class} in ORKG has a link to the corresponding resource in the Geonames knowledge graph. Querying Geonames is done via this link. According to its schema, the Geonames knowledge graph offers a variety of relations for the described resources. We are interested in the \textit{parent feature} which annotates the parent entity of any given other entity (i.e., show the hierarchy of locations in Geonames). We propose to implement the solution, using API request to find the hierarchy of the location. Getting the hierarchy, exploration at various levels of a region taxonomy can be done. Figure~\ref{rdf.png} shows a subset of RDF triples from the Geonames representation of the \textit{City of Bonn} entity indicating the \textit{parent feature} as well. 
By querying the Geonames graph, the hierarchy of locations can be discovered. After obtaining this hierarchy, the information can be leveraged in a faceted search system to support searching on broader locations and thus support a form of qualitative spatial reasoning\footnote{Hierarchy of Geonames: \url{https://www.geonames.org/export/place-hierarchy.html\#hierarchy}}. Figure~\ref{geonames.png} demonstrates the workflow between the ORKG and Geonames knowledge graphs. 

For instance, if a user filters a contribution comparison for studies conducted in Europe (e.g., studies involving a European population or an ecosystem in Europe), for each paper's \textit{study location}, our system checks the (RDF) description of the study location in Geonames. After evaluating in the hierarchy, whether the location has Europe in its parent features, the location is shown as a facet. If now a user chooses this facet, the correspondingly matching contribution descriptions would be displayed in the results.
Therefore, a query for exploring paper contribution descriptions that refer to a special method of research and have specific values in a specific duration of a particular region, can easily be answered.

Figure~\ref{comparison-facets.png} depicts an example of the faceted search performed on a COVID-19 contribution comparison, which consists of 31 papers. When a filter icon is selected, a dialogue box containing the relevant values for the property appears, thus enabling the user to choose some of the candidate values. When applying a filter, the colour of the filter icon changes to be recognizable, and a tool-tip about the selected values is displayed when hovering over the filter icon. Additionally, all applied filters are indicated clearly on top of the table. The results are directly reflected on the screen.

Furthermore, the system provides the opportunity to save these configurations and the subset of retrieved data as a new comparison to the database, with a permanent URL that can be shared with other researchers and users. We provide a link to the system to enable independent testing and investigation.\footnote{\url{https://www.orkg.org/orkg/comparisons}}\\
The code of the system is publicly available and documented on GitLab.\footnote{ \url{https://gitlab.com/TIBHannover/orkg/orkg-frontend}}

\section{Discussion}
Faceted search, as a search system, became popular with e-commerce services. During recent years, this search and exploration paradigm was increasingly used for developing scholarly knowledge databases, since it could better filter the search results and support the retrieval of more relevant data. It also improves data findability and reduces null-result searches. However, these benefits are not enough for a researcher who is looking for knowledge. We discuss next the key factors in evaluating a search system.

\textit{Precision} matters. The problem with the knowledge graphs mentioned in the related work section is that despite having a huge database, the data indicated in a paper is not searchable. Therefore, scientists mostly would not achieve an
accurate and relevant answer to their scientific queries. The key point is that, search on structured content, rather than full text, is likely to result in higher precision. However, it makes formulating queries also more complicated. 

\textit{Recall} is essential. The few knowledge graphs with structured content have limited databases and struggle to satisfy recall (e.g., limited to a particular research field and missing potentially relevant work outside the particular field). Hence, relevant answers to a query may not appear in the results. 

Moreover, facets are normally defined on the metadata of a publication. Few knowledge graphs with a limited database defined facets on the content of a paper. Also, the facets are fixed and static and have no flexibility according to the users' query.

While the Scholarly knowledge graph describes papers in a structured manner, the content of each paper is explorable to discover the accurate data related to a search. As the number of contributions described in a knowledge graph increases so does recall. 

Our faceted search system leverages a federation of knowledge graphs. That's why the facets are defined dynamically according to the users' query. So the results of a query can be narrowed down into a precise set of answers.

\subsubsection{Challenges.}
What made the problem of faceted search challenging for us are the following points:
\begin{itemize}
    \item Knowledge graphs are heterogeneous by nature. Different knowledge graphs have different structure. Thus, they are not compatible with a strict search system. Various schemas and APIs make the exploration of federated systems even harder.
    \item Completeness matters. The more complete the database is, the more data would be discovered. Unfortunately, some well-structured systems suffer from an incomplete data source~\cite{heist2020knowledge}.
    \item Each paper could be related to one or more research fields. Therefore, finding the appropriate facet according to the user’s search expression is challenging.
    \item Facets which are defined according to the data obtained from other knowledge graphs e.g., \textit{location}  facets, could be defined on two different occasions. The first one was during the search process. We could run an API request when a user searches for a location. The advantage of this approach is that the data is current and there is no need to prepare data beforehand. However, the disadvantage is the increase in the response time and the fragility in regard to network connectivity and service availability. The second option is to cache data from the second knowledge graph to allow for faster processing. An important advantage of this approach is better performance. We propose to implement the first approach not to cache unnecessary data.
\end{itemize}

\section{Future Work.} For future work, we plan to evaluate the proposed approach with user study (precision and recall), in particular user friendliness. We also plan to leverage more knowledge graphs for even smarter faceted search. We suggest that smart faceting may be defined for numerous data types, e.g., taxonomies, units, space and time, and numeric ranges which we briefly discuss next. Similarly to the approach described here with Geonames locations, for taxonomic data more generally we can leverage corresponding knowledge graphs to obtain hierarchies, e.g., about species, materials, chemicals, ecosystems, language, etc.

Also we plan to integrate a smart unit conversion. For example, if the user is looking for the data in \textit{meter} and the data in the knowledge graph is defined in \textit{kilometre}, an automatic conversion would be applied before processing and displaying the results.

Our focus here was on demonstrating how knowledge graphs can be leveraged to improve faceted search for the special case of qualitative spatial data. In future work, we will extend the approach to quantitative spatial data in order to enable users filtering by regions on a map and support quantitative spatial reasoning in faceted search. 

Of interest are also smart faceting on numeric ranges, such as \textit{Confidence Interval (CI)} or types with well-defined boundaries, such as time intervals, pH or degree Kelvin. Smart faceting is aware of such constraints and prompts users accordingly with additional functionality (e.g., filtering by duration) or warnings (e.g., if a given value is invalid such as -300 degrees Kelvin). 

Finally, we will explore applying ontologies for resolving the synonyms of the queries and defining facets according to them. For instance, if somebody is looking for the word \textit{covid}, data using synonymous terms such as \textit{corona}, \textit{covid-19}, \textit{sars-cov-2}, etc. should appear in results.

\section{Conclusion}
Nowadays, knowledge graphs are central to the successful exploitation of knowledge available as a steadily growing amount of digital data on the web. Such technologies are essential to lift traditional search systems from a keyword search to smart knowledge retrieval, which is crucial for obtaining the most relevant answers for a user query, especially in digital libraries. Despite improvements of scholarly search engines, traditional full-text search remains ineffective in many use cases. In this paper, we demonstrate a methodology for developing a faceted search system leveraging a federation of scholarly knowledge graphs. This search system can dynamically integrate content from further remote knowledge graphs to achieve a higher order of exploration usability on scholarly content, which can be matched and filtered to better satisfy user information needs. In future work, we will implement better support for various taxonomies and data types. In addition, we will work on integration query expansion features for discovering abbreviations and synonyms of terms in a query to further improve dynamic faceted search. 

\subsubsection*{Acknowledgements}
This work was co-funded by the European Research Council for the project ScienceGRAPH (Grant agreement ID: 819536) and the TIB Leibniz Information Centre for Science and Technology.
The authors would like to thank Mohamad Yaser Jaradeh for helpful comments.
 \bibliographystyle{splncs04}
 \bibliography{references}

\end{document}